\documentclass[twocolumn,showpacs,preprintnumbers,amsmath,amssymb]{revtex4}
\usepackage{graphicx}
\usepackage{dcolumn} 
\usepackage{bm}    
\begin{document}
\title{Threshold of transverse mode coupling instability \\
with arbitrary space charge}-
\author{V. Balbekov}
\affiliation {Fermi National Accelerator Laboratory\\
P.O. Box 500, Batavia, Illinois 60510}
\email{balbekov@fnal.gov} 
\date{\today}

\begin{abstract}
Threshold of the transverse mode coupling instability is calculated 
in frameworks of the square well model at arbitrary value of space charge 
tune shift.	
A new method of calculation is developed beyond the traditional expansion 
technique. 
The square, resistive, and exponential wakes are investigated.
It is shown that the instability threshold goes up without limit 
when the tune shift increases.
A comparison with conventional case of the parabolic potential well is 
performed.
\end{abstract}

\maketitle

\section{INTRODUCTION}
Transverse mode coupling instability (TMCI) of a bunch with space charge (SC) 
was considered in papers \cite{Bl1}-\cite{Ba3}. 
The authors have represented about similar results concerning the SC impact 
on the TMCI at a moderate ratio of the SC tune shift to the synchrotron tune 
$(\Delta Q/Q_s)$.
It follows from these papers that threshold of the instability caused by a 
negative wake increases when the ratio goes up from zero to several tens, 
at least.

However, more confusing picture appears at larger (but realistic) values of 
 this ratio like a hundred or over it.
It has been suggested in Ref.~\cite{Bu1} that the threshold growth ceases 
 above this border coming to 0 at $\,\Delta Q/Q_s\rightarrow\infty$.  
By contrast, it was asserted in Ref.~\cite{Ba1} that negative wake cannot 
 excite the TMCI in this limiting case. 

It should be noted in this regard that approximate methods of solution were 
 applied in all quoted articles.
Typically, expansion of the bunch offset in terms of some set of basic vectors,
 with subsequent truncation of the series, has been used at the modest tune 
 shift. 
However, as it has been shown later in Ref.~\cite{Ba3}, number of the 
 equations should be about proportional to $\,\Delta Q/Q_s$ to provide 
 appropriate convergence of the results with negative wake. 
Therefore, a smooth asymptotic transition is really impossible in frameworks 
 of this method.
In contrast with it, the convergence is very well with positive wake allowing 
 to reach the result by using the three-mode approximation 
 \cite{Ba2},\cite{Ba3}. 

The method which is developed in presented article does not use the expansion 
 at all, and allows to get the TMCI thresholds as a smooth function of 
 arbitrary SC tune shift. 
In the beginning of the paper, it is applied to the square potential well 
 model.
Such a model has been used earlier together with the expansion technique to 
 investigate the TMCI at modest space charge \cite{Bl1},\cite{Bu1}.
More wide results are represented in this article.
In particular, it is shown that the TMCI threshold with negative wake is about
 proportional to $\,\Delta Q/Q_s$.
It is shown as well that the conclusion still stands with resistive and 
 exponential wakes.

\section{Physical model}

\subsection{General relations}

Chromaticity will not be considered in this paper because it is a factor of 
 small importance for the TMCI  \cite{Ba2}.
Then the transverse coherent displacement of a bunch in the rest frame can be 
 represented as the real part of the function
\begin{eqnarray}
 X(\theta,u,t) = Y(\theta,u)\exp\big[-iQ_\beta(\theta+\Omega_0t)
-i\Omega_0\nu t\big] 
\end{eqnarray}
 where $\,\theta\,$ and $\,u\,$ are coordinate and momentum of a particle in 
 the longitudinal phase space, $\,t\,$ is time, $\,Q_\beta$ and $\,\Omega_0$ 
 are the central betatron tune and the revolution frequency, $\,\nu$ is an 
 addition to the tune due to the wake field.
The last will be characterized in the paper by the function $\,q(\theta)$. 
Then the function $\,Y$ satisfies the equation 
\begin{eqnarray}
 \nu Y +iQ_s\frac{\partial Y}{\partial\phi}+\Delta Q(Y-\bar Y) = \nonumber \\ 
 2\int_\theta^{\infty}q(\theta'-\theta)\bar Y(\theta')\rho(\theta')\,d\theta' 
\end{eqnarray}
 where $\,\phi\,$ and $\,Q_s$ are phase and tune of the synchrotron 
 oscillations, $\,\Delta Q\,$ is the space charge produced betatron tune shift
 \cite{Ba1}.
The variable $\,\bar Y$ is defined by the relations 
\begin{subequations}
\begin{eqnarray}
\rho(\theta)\bar Y(\theta)=\int_{-\infty}^\infty \Phi(\theta,u)Y(\theta,u)\,du,
\\  \rho(\theta) = \int_{-\infty}^\infty \Phi(\theta,u)\,du
\end{eqnarray}
\end{subequations}
 with $\,\Phi$ as the normalized distribution function of the bunch.

The function $\,q(\theta)$ is proportional to the transverse wake potential 
 $\,W_1(z)$ where  $\,z=R\theta$  and $\,R$ is the machine radius:
\begin{eqnarray}
q(\theta) = \frac{r_0N_bRW_1(-R\theta)}{8\pi\beta\gamma Q_\beta} 
\end{eqnarray}
with $\,r_0=e^2/mc^2$ as the particle electromagnetic radius, $\,\beta$ and 
 $\,\gamma$ as its normalized velocity and energy, and $\,N_b$ as the bunch 
 population \cite{Ba2}.
It is convenient to split the function into 2 multipliers characterizing the wake normalized amplitude and its shape: $\,q(\theta) = q_0 w(\theta)$ where
\begin{subequations}
\begin{eqnarray}
 q_0=2\int_{-\infty}^\infty\rho(\theta)\,d\theta\int_\theta^\infty 
 q(\theta'-\theta)\rho(\theta')\,d\theta', \\
 1=2 \int_{-\infty}^\infty \rho(\theta)\,d\theta\int_\theta^\infty 
 w(\theta'-\theta)\rho(\theta')\,d\theta'.
\end{eqnarray}
\end{subequations}
With this definition, tune of the lowest (rigid) bunch mode is 
\begin{eqnarray}
 \nu_{rigid}=q_0 \qquad {\rm at} \qquad Y=\bar Y = 1.
\end{eqnarray}
Of course, this expression can be valid at full only at $\,q_0\ll Q_s$ when 
 coupling of the modes is negligible.
However, it does not matter in this consideration because the value is used 
merely for the  normalization.

Separating even and odd parts of the function $\,Y=Y^{+}(\phi)+Y^{-}(\phi)$ 
 one can get the equation
\begin{eqnarray}
 \hat\nu Y^{+}+Q_s^2\frac{\partial}{\partial\phi} \left(\frac{\partial 
 Y^{+}}{\hat\nu\,\partial\phi} \right) = \Delta Q\bar Y +  \nonumber \\
 2\int_\theta^\infty q(\theta'-\theta)\,\bar  Y(\theta')\rho(\theta')\,
 d\theta', 
\end{eqnarray}
 where $\,\hat\nu(\theta)=\nu+\Delta Q(\theta)$.

\subsection{Square potential well}

In practice, it is more convenient to use a variable $\,\vartheta\propto\theta$ 
 as the longitudinal coordinate adjusted to the bunch length.
For the square well model used in the paper, the best choice is the well and 
 the bunch location in the interval $\,0<\vartheta<\pi$.
Then 
\begin{eqnarray}
 \vartheta=|\phi|, \quad \phi=\vartheta\frac{u}{|u|}, \quad 
\rho(\vartheta)=\frac{1}{\pi}\qquad{\rm at}\quad 0<\vartheta<\pi
\end{eqnarray}
Note that Eq.~(7) remains in force with new variable if the normalization 
 conditions Eq.~(5) are adjusted as well to save the validity of Eq.~(6).

Any monotonous function of $\,|u|$ can be used as the synchrotron amplitude 
 in the case.
The synchrotron tune $\,Q_s$ is the most natural and convenient choice.
Therefore, taking into account that only $\,Y^+\,$ makes the contribution 
 into Eq.~(3a), and that it is an even function of $\,u,\,$ one can rewrite 
 this expression in the form
\begin{eqnarray}
 \bar Y(\vartheta) = \int_0^\infty F(Q_s)Y^+(\vartheta,Q_s)\,dQ_s
\end{eqnarray}
 with the normalization condition
\begin{eqnarray}
 \int_0^\infty F(Q_s)\,dQ_s = 1.
\end{eqnarray}
Because $\,Y^+(\phi,Q_s)\,$ is an even and periodic function of $\,\phi$, 
 it is sufficient to consider the interval $\,0<\phi<\pi\,$ where  
 $\,\phi=\vartheta$.
Taking into account as well that $\,\Delta Q=const\,$ in the square potential 
 well, one can represent Eq.~(7) in the form
\begin{eqnarray}
 Q_s^2\frac{\partial^2Y^{+}}{\partial\vartheta^2}+\hat\nu^2 Y^{+} 
=\hat\nu\Delta Q\bar Y + \nonumber\\ \frac{2\hat\nu}{\pi}\int_\vartheta^\pi 
 q(\vartheta'-\vartheta)\bar Y(\vartheta')\,d\vartheta'.
\end{eqnarray}
The boundary conditions of the equation are
\begin{eqnarray}
 \frac{\partial Y^+}{\partial\theta}(0,Q_s) 
=\frac{\partial Y^+}{\partial\theta}(\pi,Q_s)=0
\end{eqnarray}
which relation also follows from periodicity and parity of the function 
$\,Y^+$.

\section{Hollow bunch with a square wake}

A hollow bunch will be considered below.
Its distribution function is
\begin{eqnarray}
 F(Q_s) = \delta(Q_s-Q_{s0})
\end{eqnarray}
According to Eq.~(9), $\,\bar Y(\vartheta)=Y^+(\vartheta,Q_{s0})\,$ 
 in the case, so that Eq.~(11) and  its boundary conditions  Eq.~(12) obtain 
 the form
\begin{subequations}
\begin{eqnarray}
 Q_{s0}^2\bar Y^{''}(\vartheta)+\hat\nu(\hat\nu-\Delta Q)\,\bar Y(\vartheta) 
=\nonumber\\  \frac{2\hat\nu}{\pi} \int_\vartheta^\pi q(\vartheta'-\vartheta)
 \bar Y(\vartheta')\,d\vartheta',  \\  \bar Y'(0)=\bar Y'(\pi)=0.
\end{eqnarray}
\end{subequations}
As the first step, we consider the simplest case of constant wake: 
 $\,q=q_0,\;w=1$ inside the bunch. 
It results in the equation
\begin{eqnarray}
 \bar Y^{''}(\vartheta)+\frac{\hat\nu(\hat\nu-\Delta Q)}{Q_{s0}^2}\,
 \bar Y(\vartheta) = \frac{2\hat\nu q_0}{\pi Q_{s0}^2} \int_\vartheta^\pi
 \bar Y(\vartheta')\,d\vartheta',\quad 
\end{eqnarray}
Only the case $\,q_0\le0$ will be investigated below because the positive wake 
 is an occasional and not questionable occurrence \cite{Ba2},\cite{Ba3}.

\subsection{Solution by an expansion}

This subsection pursues two goals.
First of them is a clarification of some properties of the solution for a
 further using, and second one is a comparison with the more conventional case 
 of a parabolic potential well when the expansion technique is the prevailing 
 method. 

General solution of Eq.~(15) with boundary conditions given by Eq.~(14b) 
 can be represented as the series 
\begin{eqnarray}
 \bar Y(\vartheta)=\sum_{n=0}^\infty Y_n\cos n\vartheta
\end{eqnarray}
 with unknown coefficients $\,Y_n$.
Then the equation rearranges to the form 
\begin{eqnarray}
 \sum_{n=0}^\infty \big[\,\hat\nu(\hat\nu-\Delta Q)-n^2 Q_{s0}^2\,\big]
 Y_n\cos(n\vartheta) = \nonumber\\
\frac{2q_0\hat\nu}{\pi}\sum_{n=0}^\infty Y_n   \int_\theta^\pi\cos(n\vartheta')\,d\vartheta'.
\end{eqnarray}
Multiplying this expression by $\,\cos(N\vartheta)\,$ and integrating over 
 $\,\theta$, one can obtain the series of equations for the coefficients 
 $\,Y_n$:
\begin{eqnarray}
 \big[\hat\nu(\hat\nu-\Delta Q)-N^2Q_{s0}^2\big]Y_N=\nonumber\\
 q_0\hat\nu  (2-\delta_{N,0})\sum_{n=0}^\infty R_{N,n} Y_n
\end{eqnarray}
 where $\,R_{n,n}=\delta_{n,0}$, and other elements of the $R$-matrix are
 \begin{eqnarray}
 R_{N,n}=\frac{2}{\pi^2}\int_0^\pi \cos(N\vartheta)\,d\vartheta 
 \int_\theta^\pi\cos(n\vartheta')\,d\vartheta'=\nonumber\\
 \frac{2\big[1-(-1)^{N-n}\big]}{\pi^2(N^2-n^2)}.
\end{eqnarray}
Its small fragment is represented in Table I.
\begin{table}[b]
\begin{center}
\caption{Fragment of the matrix $R_{N,n}\quad  (A=4/\pi^2$) }
\vspace{10mm}
\begin{tabular}{|c|c|c|c|c|c|c|}
\hline -
$N\rightarrow$  &~~~0~~~&~~~1~~~&~~~2~~~&~~~3~~~&~~~4~~~&~~~5~~~\\
\hline
$  ~ n=0 ~   $  &    1    &   $A$   &   0     & $A/9$ &    0   &  $A/25$ \\
$  ~ n=1 ~   $  &   $-A$  &    0    &  $A/3$  &   0   & $A/15$ &  0      \\
$  ~ n=2 ~   $  &     0   &  $-A/3$ &   0   & $ A/5$  &    0   &  $A/21$ \\
$  ~ n=3 ~   $  &  $-A/9$ &    0    & $-A/5$  &   0   & $A/7$  &  0      \\
$  ~ n=4 ~   $  &    0    & $-A/15$ &   0   & $-A/7$  &    0   &  $A/9$  \\
$  ~ n=5 ~   $  & $-A/25$ &    0    & $-A/21$ &   0   & $-A/9$ &  0      \\
\hline
\end{tabular}
\end{center}
\end{table}

The infinite series given by Eq.~(18) can be truncated by using of the 
 assumption  $\,Y_n=0\,$ at $\,n>N_{max}$.
It results in the finite set of the equations 
\begin{subequations}
\begin{eqnarray}
 \sum_{n=0}^{N_{max}} T_{N,n}\bar Y_n=0,\\
 T_{N,n}= q_0 \hat\nu (\delta_{N,0}-2)R_{N,n}+  \nonumber\\
\big[\hat\nu(\hat\nu-\Delta Q)-N^2 Q_{s0}^2\big]\delta_{N,n}
 \end{eqnarray}
\end{subequations}
A minimal set comes with $\,N_{max}=0$ and includes only the lowest (rigid) 
 head-tail mode $\,Y_0=1$.
Eq.~(20) gives in this case  $\,T_{0,0}=0$ that is  $\,\hat\nu-\Delta Q
 =\nu=q_0$ as it is required by Eq.~(6).

General resolvability condition of the series  is $\,{\rm det}\,T=0\,$ which 
 is referred in the case to the algebraic equation of power 
 $\,P=2(N_{max}+1)\,$.
It has $\,P\,$ roots which are the real numbers at $\,q_0=0$ 
 \cite{Bl1},\cite{Bu1}:
\begin{eqnarray}
 \hat\nu_{n,\pm}=\frac{\Delta Q}{2}\pm\sqrt{\frac{\Delta Q^2}{4}+n^2Q_{s0}^2},
\end{eqnarray}
 where $\,n = 0,\,1,\dots,N_{max}$.
Therefore, the following steps can be used to resolve the problem and to find 
 the TMCI threshold at arbitrary $\,\Delta Q$ and $q_0$:
\\

1. To select some values  $\,N_{max}$ and  $\,\Delta Q/Q_{s0}$.

2. To choose some trial value of $\,q_0/Q_{s0}$.

3. To find the matrix $\,T(\hat\nu/Q_{s0})\,$ and to calculate its determinant 
 with taken parameters and variable $\,\hat\nu/Q_{s0}$.

4. To find number of real roots of the equation by the count how many times 
 the determinant changes sign at increasing $\,\hat\nu$.

5. To repeat the attempts with higher value of $\,|q_0/Q_{s0}|$ until number 
 of the real roots decreases.
It will mean that a pair of complex roots appears in this point, 
 and the reached value of $\,q_0$ is just the TMCI threshold with taken SC 
 tune shift at given approximation.

6. To check the convergence of the results by comparison of the thresholds 
 obtained with different $\,N_{max}$.
\\

Some results of the calculation are presented in Fig.~1 where the TMCI
 threshold of a negative wake is plotted against the tune shift at different 
 $\,N_{max}$.
The black drop-down curve belongs to all the approximations.
It is seen that any higher approximation follows the course which the lower 
 ones have charted, and provides its continuation to higher $\,\Delta Q$.
In contrast with it, the coming back lines of different color do not repeat 
 each other so they cannot be treated as the credible results.  
It allows to conclude that, with negative wake, threshold value of $\,|q_0|\,$ 
 is an increasing function of the SC tune shift in the considered range of 
 $\,\Delta Q/Q_0$, and that rather large number of $\,N_{max}$ is needed to 
 reach the correct result with higher shift.
\begin{figure}
 \includegraphics[width=80mm]{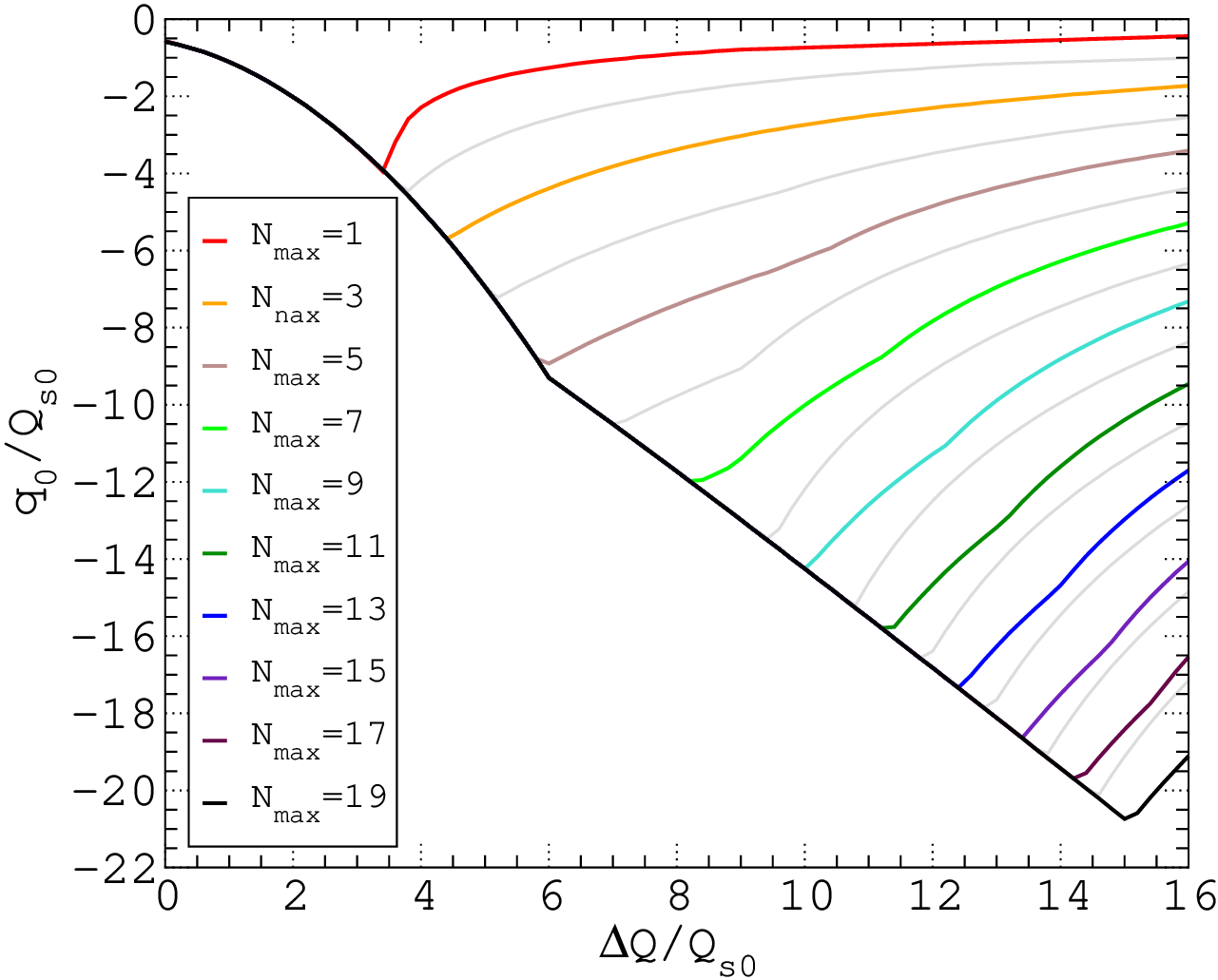}
 \caption{TMCI threshold of a hollow bunch in a square potential well against 
 SC tune shift.
Different curves are obtained with different $\,N_{\rm max} = 1,\,...\,19$.
Each of them has a restricted region of applicability which expands with 
$\,N_{\rm max}\,$ growing.
The drop-down parts of the curves merge forming the sole black line.
The rising lines do not confirm each other marking ends of the applicability
 regions.}
\end{figure}
Note that a coalescence of a pair of real roots in the $(q_0$--$\hat\nu)$ 
 plane prefaces appearance of the complex roots. 
The function $\,\hat\nu(q_0)$ satisfies the condition
\begin{eqnarray}
  \frac{d\hat\nu}{dq_0} \rightarrow \infty
\end{eqnarray}
 in the coalescence point.
The condition is valid independently on $\,N_{max}$ so it can be used to 
 identify the instability threshold with any truncation of the expansion,
 or without the expansion at all.
\begin{figure}
 \includegraphics[width=80mm]{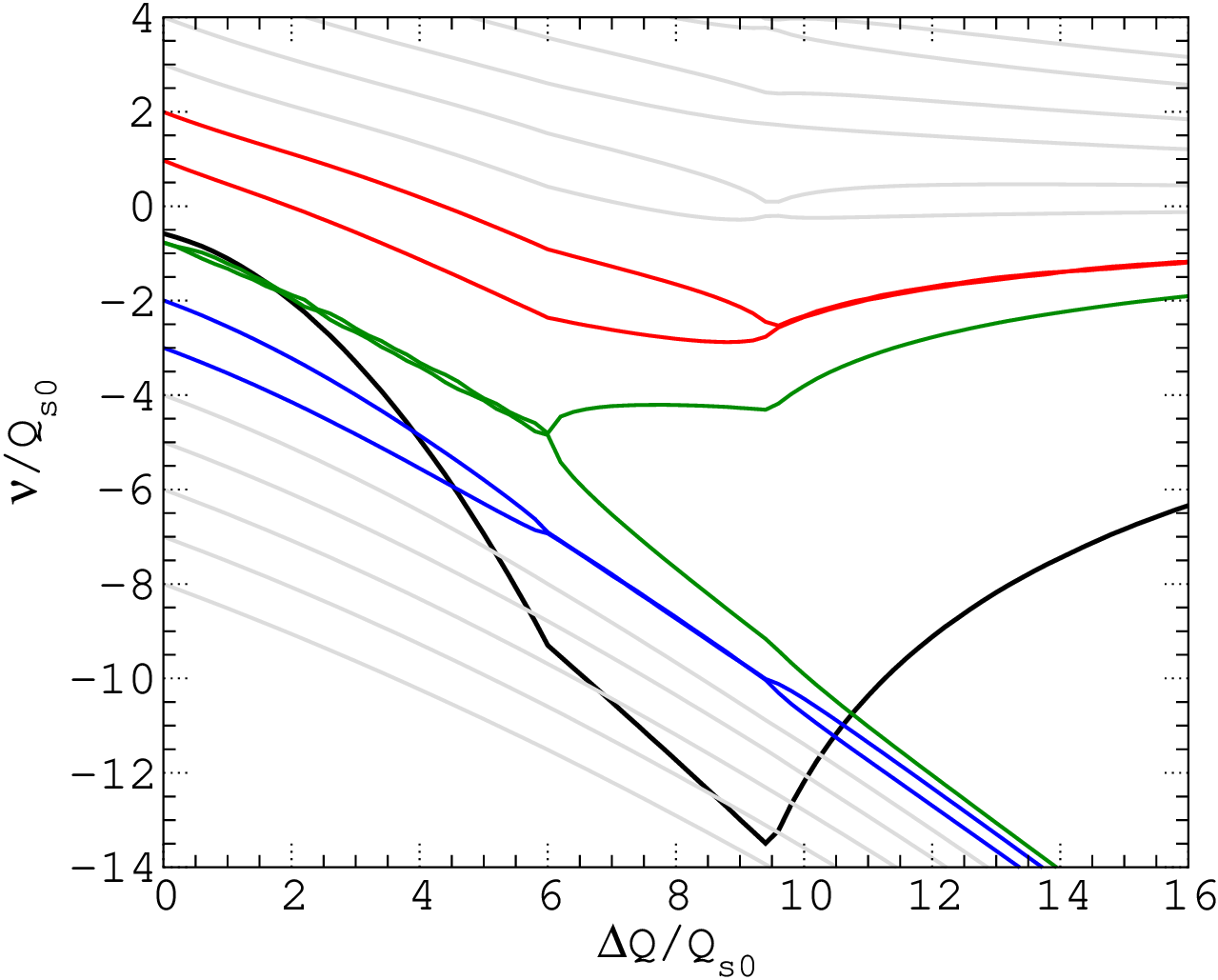}
 \caption{Threshold bunch spectrum at $\,N_{max}=8$.
The solid black line is the TMCI threshold in this approximation.
Other lines present the bunch eigentunes just before the instability appears. 
Coalescence of different multipoles is responsible for the instability at 
 different $\,\Delta Q/Q_{s0}$.
The essential lines are emphasized by colors.}
\end{figure}

An additional information is provided by Fig.~2 where complete observed bunch 
 spectrum is plotted against $\,\Delta Q/Q_{s0}\,$  at $\,N_{max}=8$.
Corresponding threshold value of $\,q_0/Q_{s0}$ is plotted in the figure 
 as well by the bold black line.
As it should be in the threshold, there is a pair of coalesced spectral lines 
 in any part of the plot.
At $\,\Delta Q/Q_{s0}<6$, they are presented as green lines which are 
 identified as the multipoles $\,m=0\,$ and $\,m=-1$.
The green lines diverge after $\,\Delta Q/Q_{s0}>6$, and the instability 
 transforms to a coalescence of the modes $\,m=-2\,$ and $\,m=-3$ 
 (the blue lines).
Finally, coalescence of the modes $\,m=1\,$ and $\,m=2\,$  appears at 
 $\,\Delta Q/Q_q>9.5\,$ which case is presented by red lines. 
However, in contrast with previous pairs, this coalescence is not confirmed 
 by calculations with higher $N_{max}$.
It allows to conclude that this coalescence is a non-physical effect appearing 
 out the region of applicability of used approximation.
The statement is confirmed by Fig.~3 where tunes of the essential modes are 
 plotted against SC tune shift at different $\,N_{max}$.
It is seen that the "responsibility" of the modes $m=-2$ and $m=-3$ for the 
 instability extends to higher values of the tune shift (blue symbols). 

It is important to emphasize a similarity of these results to those obtained 
 with the model of parabolic potential well \cite{Ba3}. 
Besides the general resembling, there is a proximity of the numerical results. 
For example, applicability region of the approximation $\,N_{max}=12\,$  
 is $\,\Delta Q/Q_{s0}<12$ in both cases, and the calculated TMCI threshold 
 is $\,q_0/Q_{s0}=-16.7$ according to Fig.1 and $-18$ according to \cite{Ba3}.
It occurs in spite of the fact that the parabolic bunch has much richer 
 spectrum than the square one, due to the higher radial modes which are 
 absent in the last case.
However, it has been shown in Ref.~\cite{Ba3} that only lowest radial modes 
 are capable to coalesce producing the TMCI. 
The square well model represents this part of the spectrum rather correctly 
 to calculate the valid TMCI threshold. 
\begin{figure}
 \includegraphics[width=75mm]{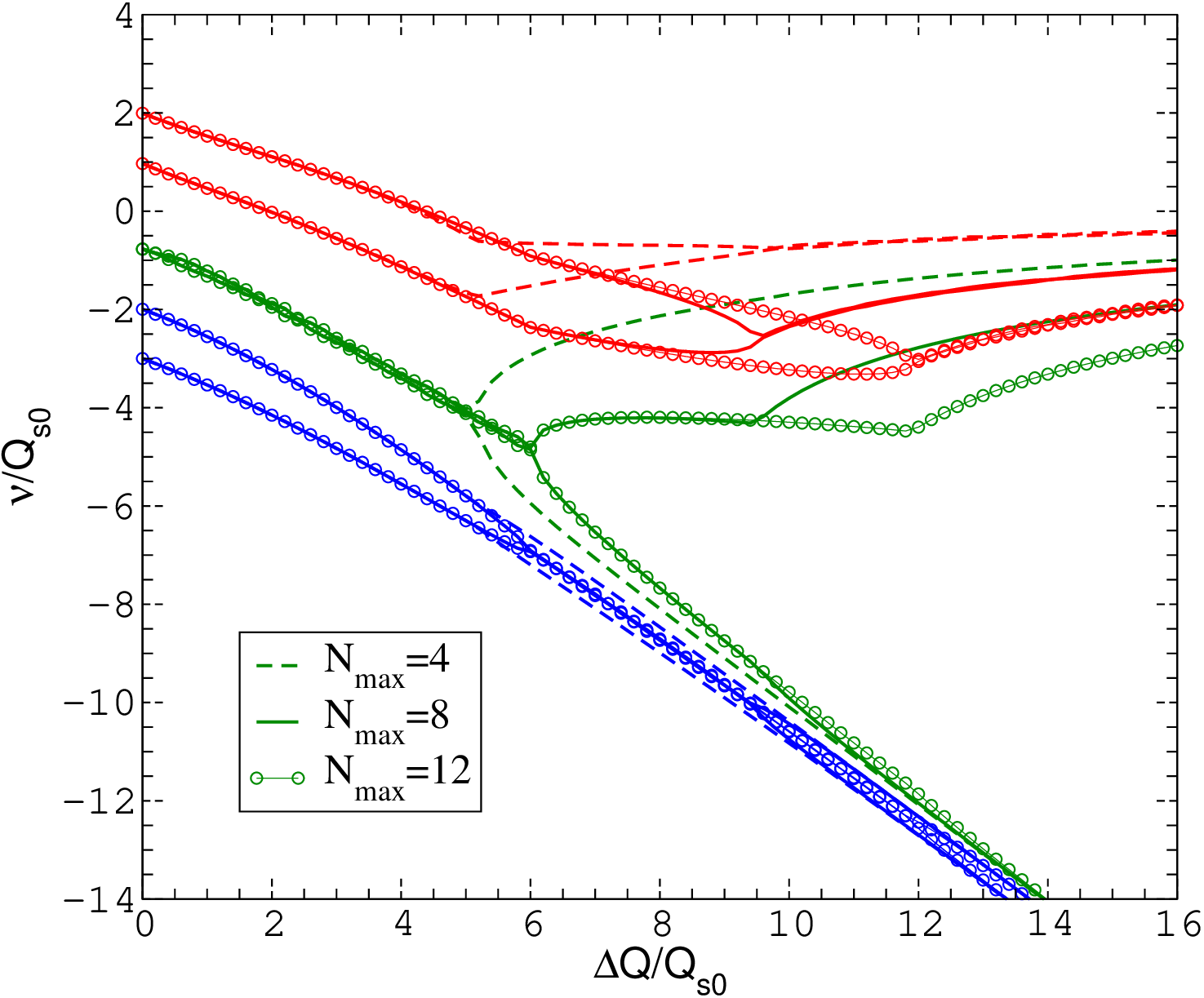}
 \caption{The threshold bunch spectrum obtained with different $\,N_{\rm max}$.
 Only essential (capable to coalesce) spectral lines are shown.}
\end{figure}

\subsection{Solution without expansion.}

Another method of solution of Eq.~(15) which is considered in this subsection 
 is free from additional assumptions and therefore is usable with any value of 
 $\,\Delta Q/Q_{s0}$.
We will use the notation
\begin{eqnarray}
 {\cal P}=\frac{\hat\nu(\hat\nu-\Delta Q)}{Q_{s0}^2}, 
 \qquad {\cal Q}=\frac{q_0\hat\nu}{Q_{s0}^2}
\end{eqnarray}
 to rewrite Eq.~(15) in the form
\begin{eqnarray}
 \bar Y''(\vartheta)+{\cal P}\bar Y(\vartheta)=
\frac{2{\cal Q}}{\pi} \int_\theta^\pi \bar Y(\vartheta')d\vartheta'
\end{eqnarray}
 which can be reduced to the proper differential equation
\begin{eqnarray}
 \bar Y'''+{\cal P}\bar Y'+\frac{2{\cal Q}}{\pi}\bar Y=0.
\end{eqnarray}
Similar equation has been investigated in Ref.~\cite{Ba2}. 
It follows from the paper that, at any real $\,\cal Q$, there is an infinite 
 discrete set of the eigenfunctions $\,\bar Y^{(k)}$ with real eigennumbers 
 $\,{\cal P}^{(k)}$ which satisfy the equation.
Actually the equation is being solved in this paper step by step with 
 arbitrary value of $\,{\cal Q}\,$ and some trial value of $\,{\cal P}$, 
 using the initial conditions 
\begin{equation}
\bar Y(\pi)=1,\qquad  \bar Y'(\pi)=0, \qquad \bar Y''(\pi)=-{\cal P}
\end{equation}
 and coming back to the point $\,\vartheta=0$. 
The values of $\,{\cal P}\,$ assuring the condition $\,\bar Y'(0)=0\,$ have 
 to be separated as the valid eigenvalues. 
Some of them are plotted in Fig.~4.

 \begin{figure}
 \includegraphics[width=80mm]{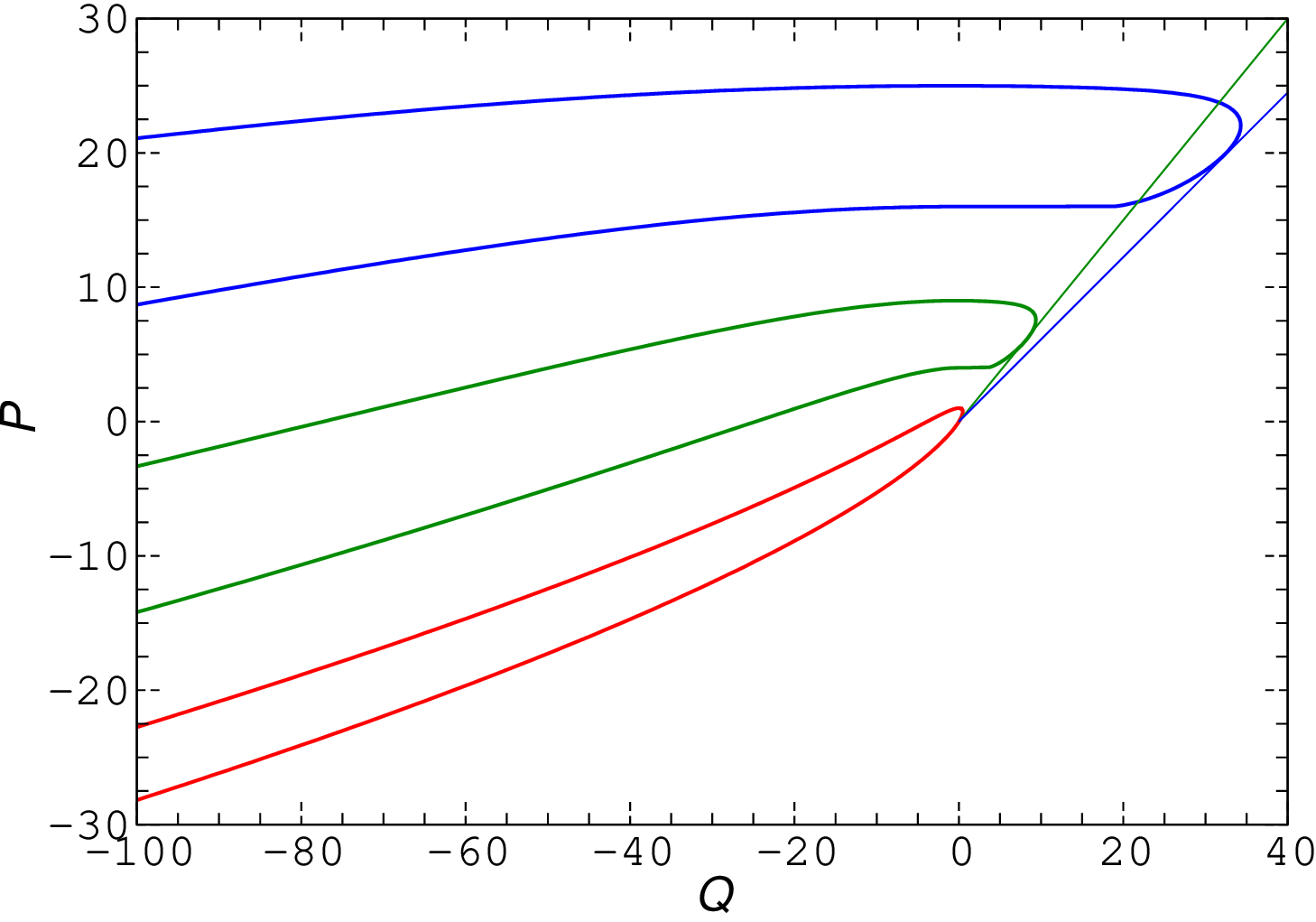}
 \caption{Lowest real eigennumbers of Eq.~(24) and (25).
Thin straight lines are tangent to the curves of corresponding color.} 
 \end{figure}
Obtained function $\,{\cal P(Q)}\,$ has to be imaged into the plane 
$\,(\hat\nu,q_0)\,$ applying the transformations 
\begin{eqnarray}
 \hat\nu = \frac{\Delta Q}{2}\pm\sqrt{\frac{\Delta Q^2}{4}+{\cal P}Q_0^2}, 
 \qquad q_0 = \frac{Q_0^2{\cal Q}}{\hat\nu}
\end{eqnarray}
 which follow from Eq.~(23).
Any point of the family represents a real eigentune of the bunch at taken SC 
 tune shift.
They form several lines representing tunes of different head-tail modes
 some of which are being shown in Fig.~5. 
These modes are stable at  rather small value of $\,q_0$  because chromaticity 
 is not included in the consideration. 
The instability can arise due to coalescence of some neighboring lines  at 
 rather large wake field.
According to Eq.~(22), the condition $\,dq_0=0\,$ marks the border point 
 where the TMCI threshold appears. 
\begin{figure}
 \includegraphics[width=73mm]{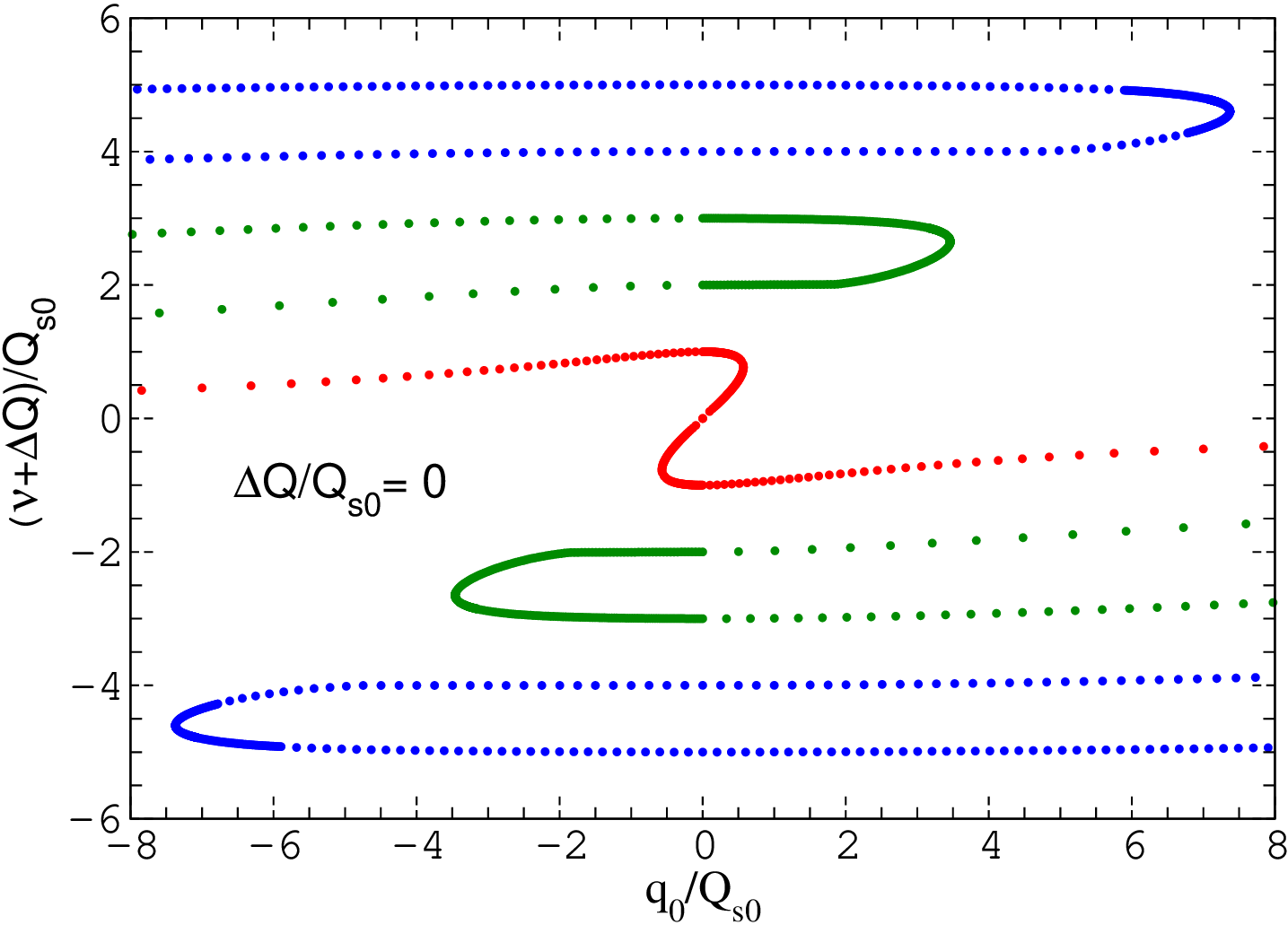}
 \includegraphics[width=73mm]{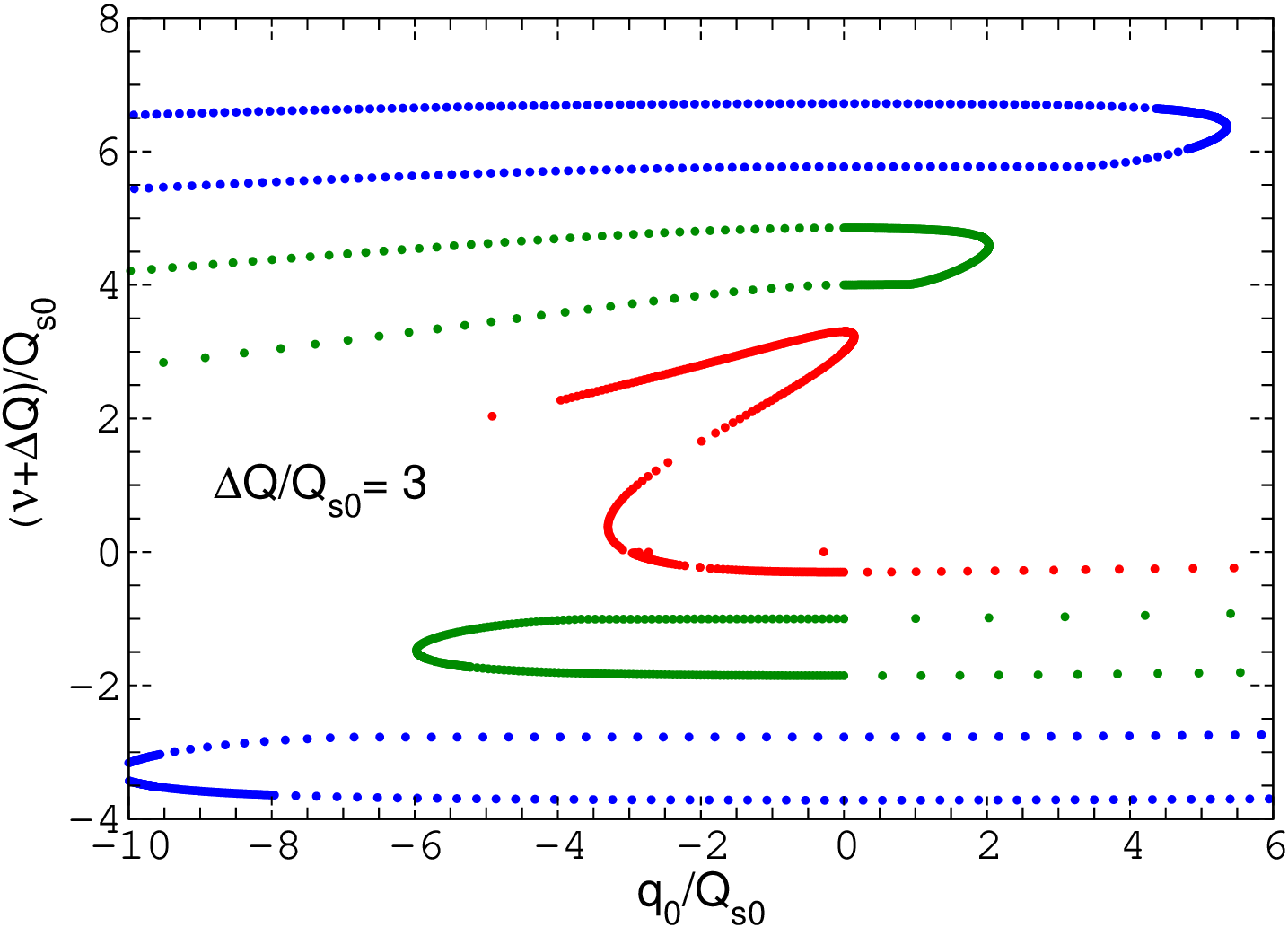}
 \includegraphics[width=73mm]{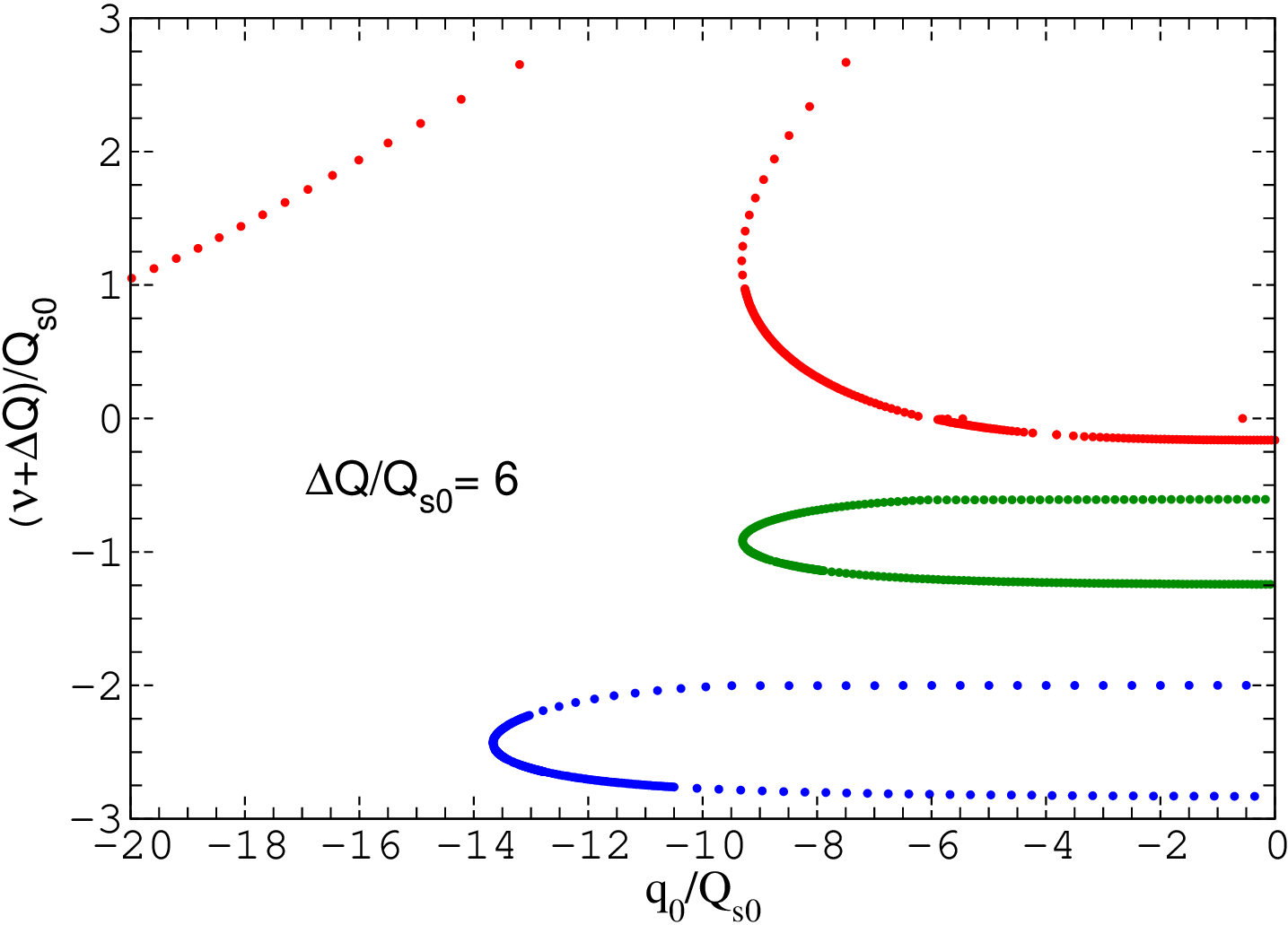} 
\includegraphics[width=73mm]{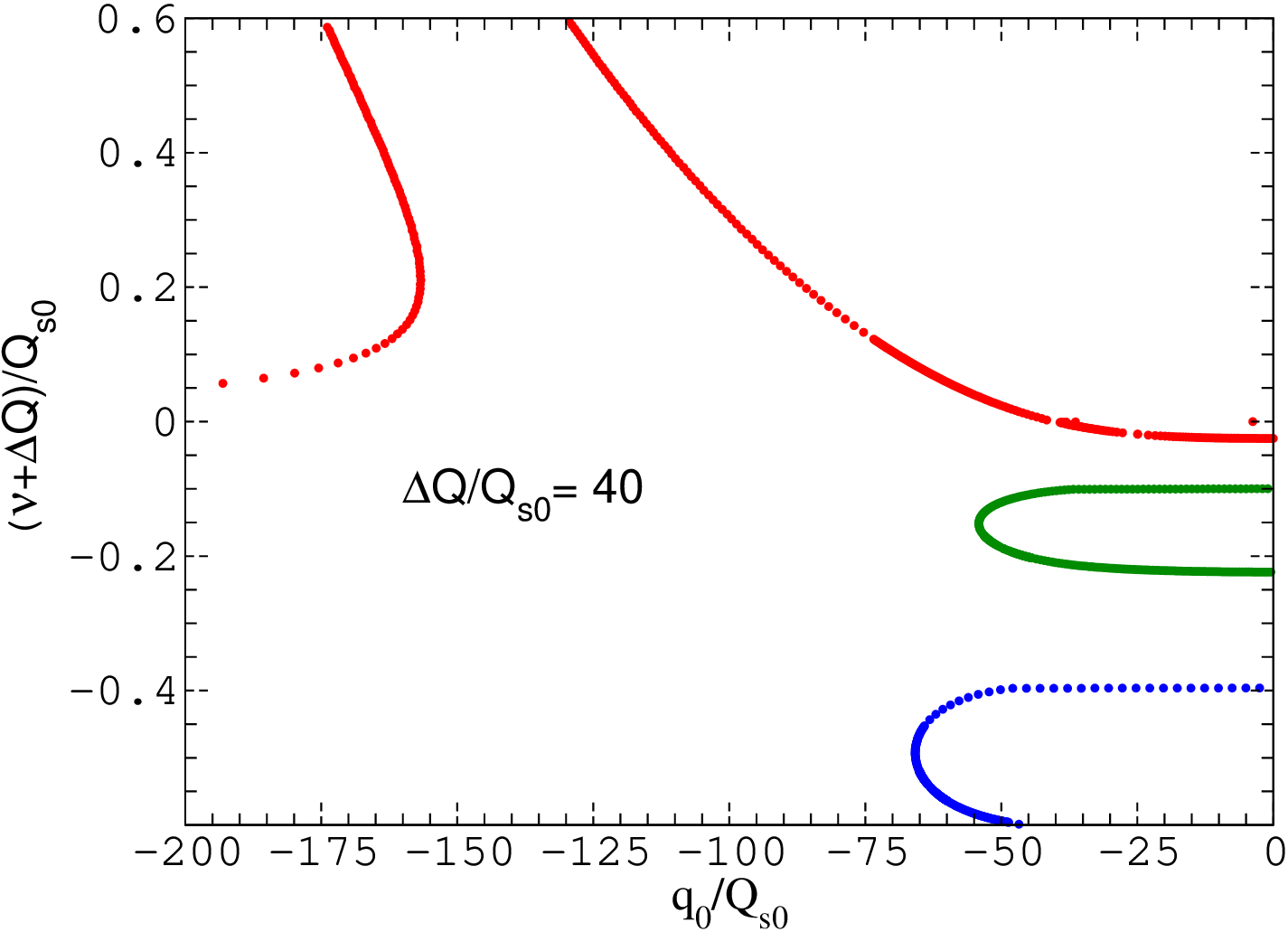}
\caption{The bunch head-tail modes against the wake strength at different 
 SC tune shifts. 
The extremal points where $dq_0=0$ mark the beginning of the instability 
 region.
 Only negative wakes are displayed at $\,\Delta Q/Q_{s0} \ge 6$}. 
 \end{figure}

 \begin{figure}
 \includegraphics[width=85mm]{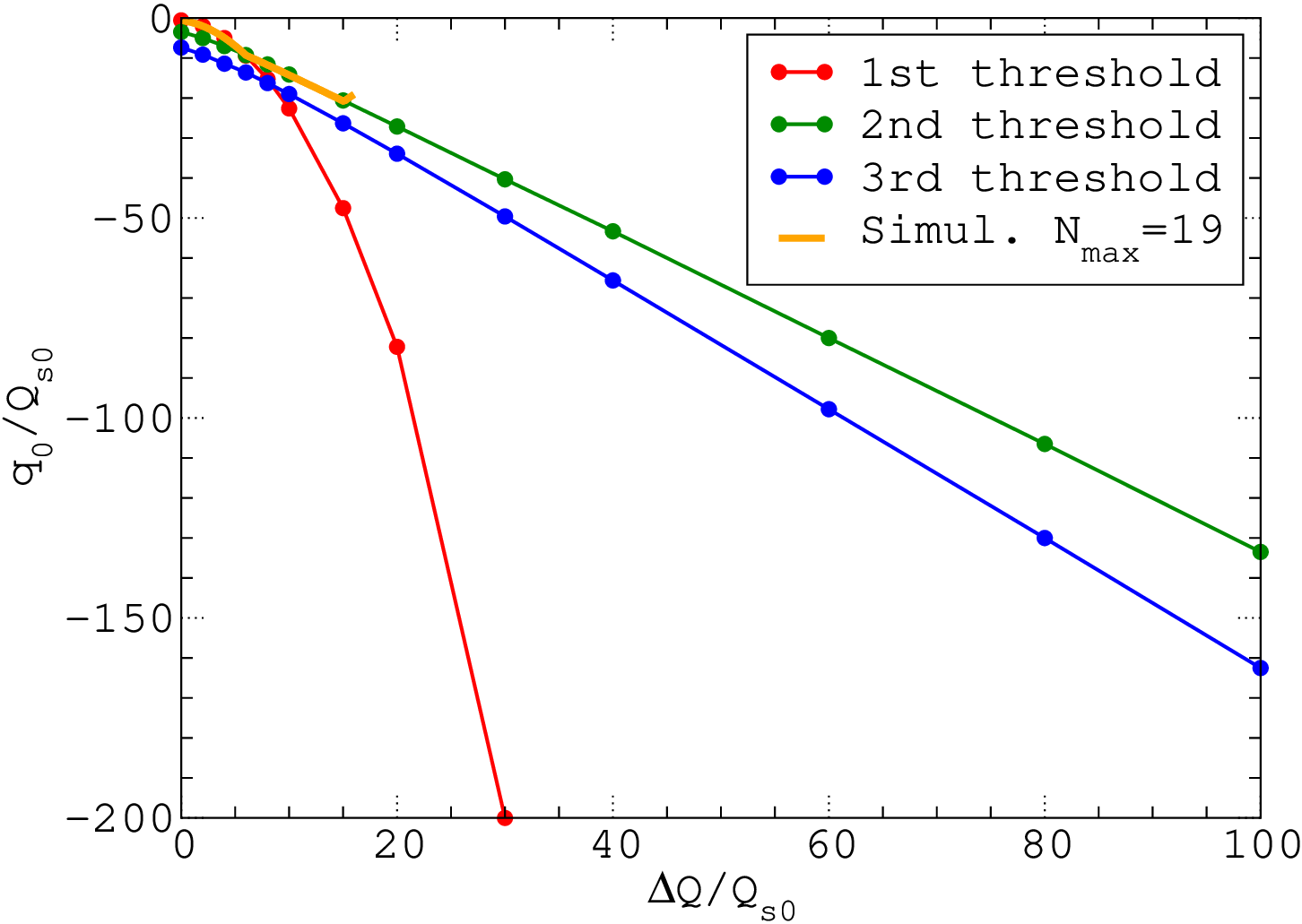}
 \caption{Instability threshold of the lowest TMCI modes.} 
 \end{figure}
All the tune lines have a well known form at $\,\Delta Q=0$ 
 (see e.g. \cite{Ng1}).
Without wake, the eigentunes form set of the multipoles $\nu_m=mQ_{s0}$.
Some of them can coalesce at higher $\,q_0$ marking a beginning  of the 
 instability region.
Corresponding threshold values of several low TMCI modes are at $\,\Delta Q=0$ 
$$
 (q_0/\Omega_0)_{thresh}\quad = \quad\pm 0.567, \quad \pm 3.46, \quad \pm 7.37
$$
 which result agrees with Ref.~\cite{Ba2}.
Other graphs of Fig.~5 illustrate deformation  of the tune lines, and movement
 of the threshold points because of  the space charge impact.
The picture is very simple with positive wake when the thresholds of all 
 unstable modes monotonously decrease tending to 0 at 
 $\Delta Q\rightarrow\infty$ \cite{Ba2}. 
Therefore only the cases $q_0<0$ are plotted in Fig.~5 at 
 $\,\Delta Q/Q_{s0}\ge 6$ and are commented below. 

It is seen that the threshold of any unstable mode increases in modulus at 
 increasing $\,\Delta Q$.
However, different modes have different velocity of the movement.
The mode caused by coalescence of the multipoles $\,m=0$ and $\,m=-1$ is the 
 most unstable at $\,\Delta Q/Q_0<6$.
However, threshold of this mode (further marked as $\,M_{0,-1}$) rather 
 rapidly raises with $\,\Delta Q$ moving to the left and yielding the role 
 of the most unstable mode to $\,M_{-2,-3}$ at $\,\Delta Q/Q_0=6$.
Next mode $M_{-4,-5}$ is more stable at any space charge.

Obtained thresholds of these coupled modes are plotted against the SC tune 
 shift in Fig.~6.
Results of the expansion technique with $\,N_{max}=19\,$ are added to the plot 
 being shown by the bold orange line. 
There is a perfect coincidence of the results at $\,\Delta Q/Q_0\le 15$ that 
 is in the applicability region of the expansion technique,
 as it has been specified above.
 
It follows from Fig.~5 that the tunes of the potentially unstable modes 
 satisfy the condition  $\,|\hat\nu|\ll\Delta Q$ at $\,\Delta Q/Q_{s0}\gg1$.
One can see that similar result can occurs only under the condition 
 $|{\cal P}| \ll \Delta Q/Q_{s0}$ when Eq.~(27) obtains the form 
\begin{equation}
\hat\nu \simeq -\frac{{\cal P}Q_{s0}^2}{\Delta Q},
\qquad\qquad  q_0 \simeq -{\frac{\cal Q}{\cal P}} \Delta Q
\end{equation}
According to these equations, the value of $\,q_0$ varies when the point 
 $({\cal Q,P})$ moves along one of the curves in Fig.~4.
The TMCI threshold obtained by Eq.~(22) appears in the point where the 
 condition  $\,dq_0=0$ is fulfilled, that is $\,d{\cal P/P}=d\cal Q/Q$.
It is the point of tangency of the curve with the straight line 
 $\,{\cal P}=k{\cal Q}$ where $k$ is a constant.
These tangents are shown in Fig.~4 by  green and blue straight lines 
 providing $\,k_g=0.75$ and $\,k_b=0.61$.
Therefore the asymptotic TMCI thresholds of corresponding modes are: 
\begin{equation}
 (q_0)_g=-1.33\, \Delta Q,\qquad\qquad (q_0)_b=-1.63\,\Delta Q. 
\end{equation}
 in agreement with Fig.~6.
Note that the procedure is unfit for the red line resulting in $\,k_r=0/0$, 
 because $\,\hat\nu\nrightarrow 0$ in the case.

\section{Resistive wall wake.}

Resistive wall is the most common and important source of transverse 
 instability in circular accelerators. 
Its wake function reaches a maximum at the distance $\,z=b/\gamma$ from 
 the source with $\,b$ as the beam pipe radius.
If the bunch length satisfies the condition $\,z_b\gg b/\gamma$, 
 and the wall is thick, the simplest relation for the transverse wake function
 is applicable:
\begin{equation}
 W_1(z)=-\frac{4R}{b^3}\sqrt{\frac{c}{\sigma|z|}}
\end{equation}
 where $\,\sigma$ the pipe wall conductivity (see e.g. \cite{Ng1}). 
According to Eqs.~(4) and (5), corresponding basic tune shift is:
\begin{equation}
 \nu_{rigid}=q_0=-\frac{4r_0R^2N_b}{3\pi\beta\gamma b^3Q_c}
 \sqrt{\frac{c}{\sigma z_b}}.
\end{equation}
Therefore, in agreement with the accepted conditions, the normalized wake 
 function in Eq.~(11) is
\begin{equation}
 q(\vartheta'-\vartheta) = \frac{q_0\kappa}{\sqrt{\vartheta'-\vartheta}}, 
\qquad \kappa = \frac{3\sqrt{\pi}}{8} 
\end{equation}
Instability of similar bunch was considered in Ref.~\cite{Ba3} using the 
 expansion technique at $\,\Delta Q/Q_{s0}\le 9$. 
Now we will investigate the problem without this restriction using equation 
 like Eq.~(24) which provides in the case
\begin{subequations}
\begin{eqnarray}
 \bar Y''(\vartheta)+{\cal P}\bar Y(\vartheta)=\frac{2{\cal Q\kappa}}{\pi} 
 \int_\vartheta^\pi\frac{\bar Y(\vartheta')\,d\vartheta'}{\sqrt{\vartheta'-\vartheta}},\\
 \bar Y'(0)=\bar Y'(\pi)=0,
\end{eqnarray}
\end{subequations}
 \begin{figure}
 \includegraphics[width=83mm]{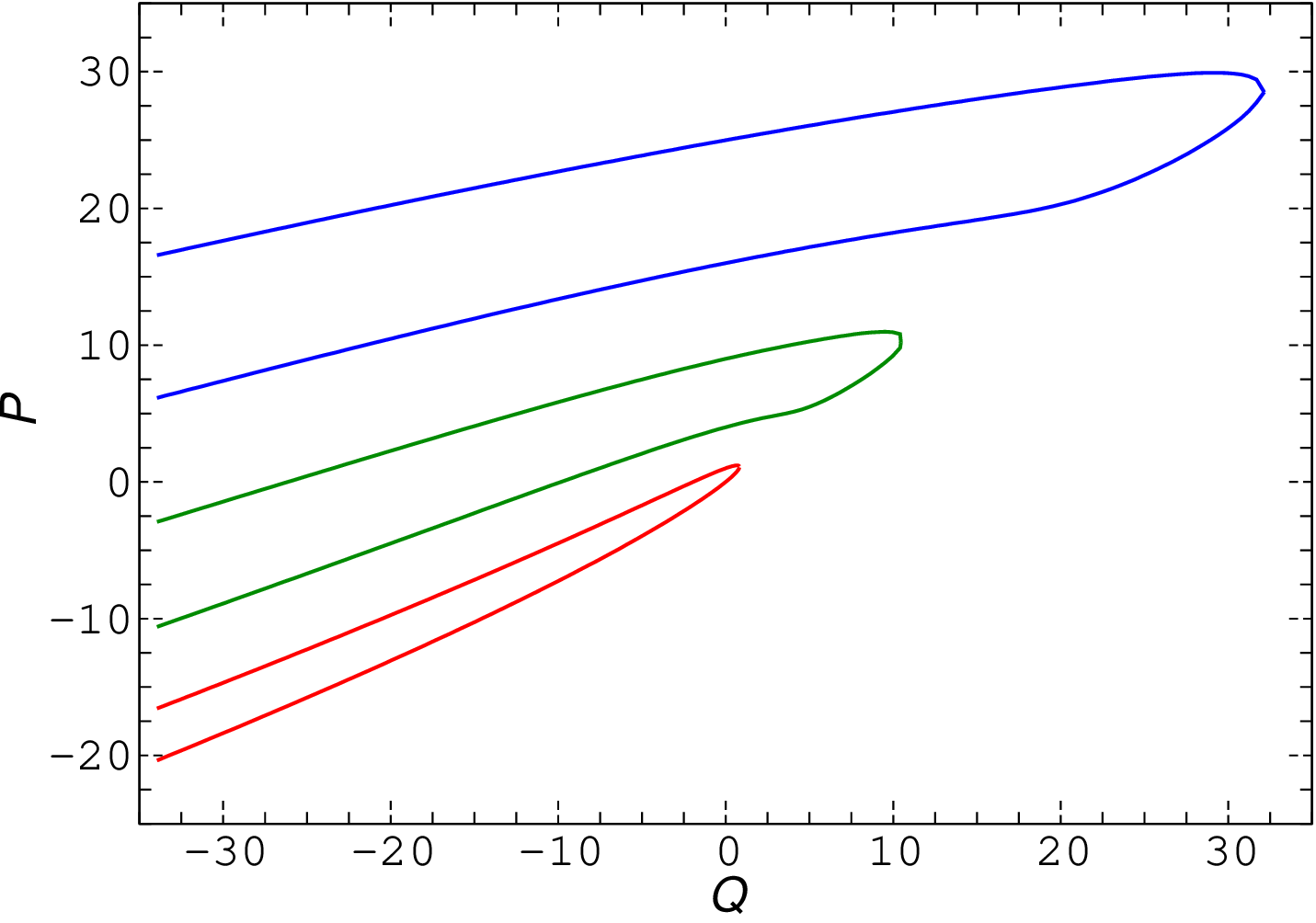}
 \caption{Lowest real eigennumbers of the resistive wall wake, Eq.~(33).} 
 \end{figure}
In contrast with Eq.~(24), this equation is not reducible to the pure 
 differential form like Eq.~(25). 
Nevertheless, the step-by-step method of the solution is applicable as above 
 being enhanced by calculation of the integral. 
The result is represented in Fig.~7 where six lower eigennumbers of the 
 equation are shown.  

These curves are so similar to those in Fig.~4 that there is no need to plot 
 the tune lines like Fig.~5.
The statement pertain equally to the possibility to determine the asymptotic 
 behavior of the threshold by the build-up of a tangent to the green line. 
Therefore we represent in Fig.~8 only the net result that is the TMCI 
 threshold with the resistive wall wake against the SC tune shift.
At $\,\Delta Q=0$, the threshold is a little larger in comparison with the 
 square wake: $\,(q_0)_{thresh}/Q_{s0}\simeq -0.90$ instead of $-0.57$. 
However, henceforth it grows  slower having asymptotically 
 $\,(q_0)_{thresh}\simeq -1.1\Delta Q$ instead of $-1.33\Delta Q$.

 \begin{figure}
 \includegraphics[width=87mm]{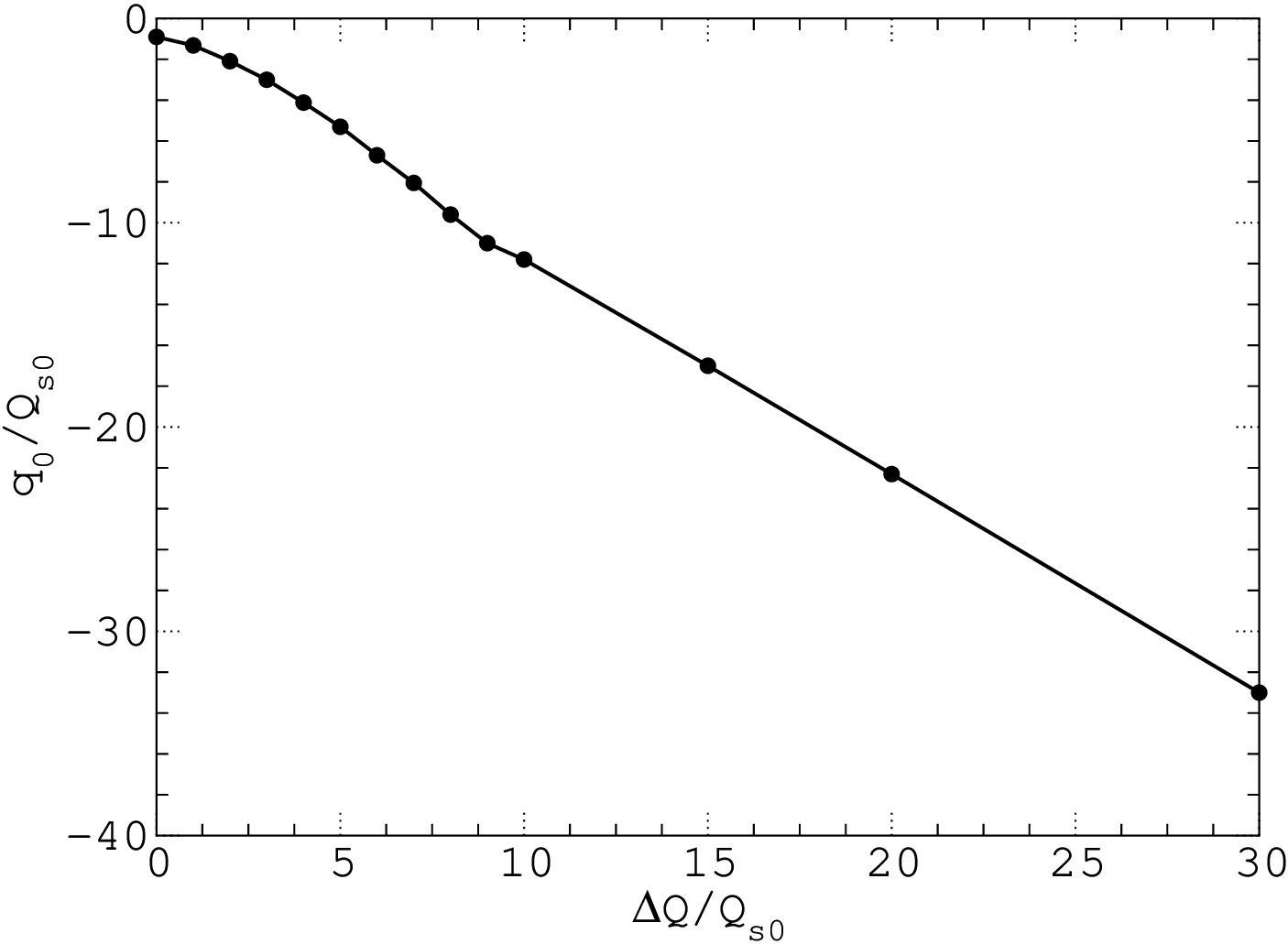}
 \caption{TMCI threshold of the resistive wall wake.
Its asymptotic value is: $\,q_0=-1.1\Delta Q$.} 
 \end{figure}

It is necessary to take into account that the resistive wake falls rather 
 slowly so it can reach the neighboring bunch ot turn and provoke a 
 multibunch/multiturn instability. 
The problem was considered in Ref.~\cite{Ba2} leading to the conclusion 
 that the TMCI effect prevails at the condition
$$
2\left(h-\frac{0.35}{\sqrt{h}}\right)\sqrt{\frac{\sigma_z}{2\pi R}} <1
$$
where $\,h$ is number of bunches, and $\,\sigma_z$ is their rms length.

\section{Exponential wake}

 \begin{figure}[b]
 \includegraphics[width=85mm]{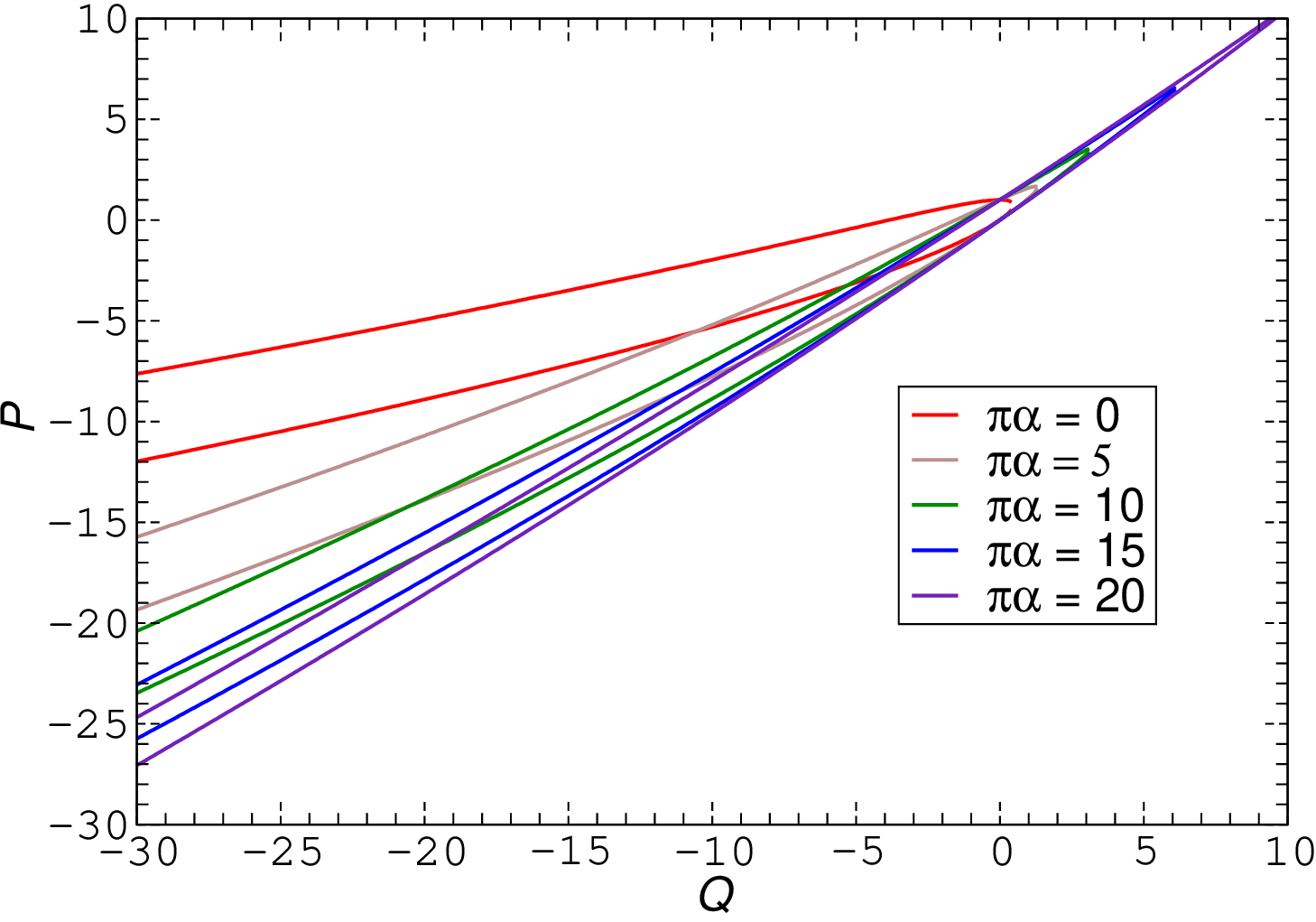}
 \caption{The lowest real eigennumbers of Eq.~(30) at different~$\alpha$. } 
 \end{figure}
 
Exponential wake of the form
\begin{subequations}
\begin{eqnarray} 
 q(\vartheta) = \kappa q_0\exp(-\alpha\vartheta), \\
 \kappa= \frac{\pi\alpha}{2} \left[1-\frac{1-\exp(-\pi\alpha)}{\pi\alpha}
 \right]^{-1}
\end{eqnarray}
\end{subequations}
 is considered in this section.
Coefficient $\,\kappa$ is added to meet the requirement  $\,\nu_{rigid}=q_0$ 
 with any $\,\alpha$. 
It is assumed as well that the wake decays rather fast after the bunch end, 
 so that it cannot reach the following bunch or turn. 

Substitution of this expression into Eq.~(11) results in the equation
\begin{eqnarray}
 \bar Y''(\theta)+{\cal P}\bar Y(\theta) = \nonumber\\
\frac{2\kappa{\cal Q}}{\pi}\exp(\alpha\theta)
\int_\theta^\pi\bar Y(\theta')\exp(-\alpha\theta')\,d\theta'
\end{eqnarray}
 where the notations embedded by Eq.~(23) are used.
It can be reduced to the proper differential form like Eq.~(25)
 with the boundary conditions like Eq.~(26)
\begin{subequations}
\begin{eqnarray}
\bar Y'''-\alpha\bar Y''+{\cal P}\bar Y'
+\left( \frac{2\kappa {\cal Q}}{\pi} -\alpha{\cal P} \right)\bar Y=0, \\
 \bar Y(\pi) = 1, \; \bar Y'(\pi) = 0, \; \bar Y''(\pi)=-{\cal P},\; \bar Y'(0) = 0.
\end{eqnarray}
\end{subequations}
Several solutions of this equation are shown in Figs.~9 and Fig.~10 at  
 $\;\pi\alpha = 0,\,5,\,10,\,15,\,20$.
The lowest eigennumbers which are represented in Fig.~9 are the analogues of the red 
 lines in Fig.~4.
The higher eigennumbers are plotted in Fig.~10 in the restricted region $\,{\cal Q}>0$.
 where instability of  higher modes can have a start, as it has been shown in Sec.~III-B.

 \begin{figure}
 \includegraphics[width=83mm]{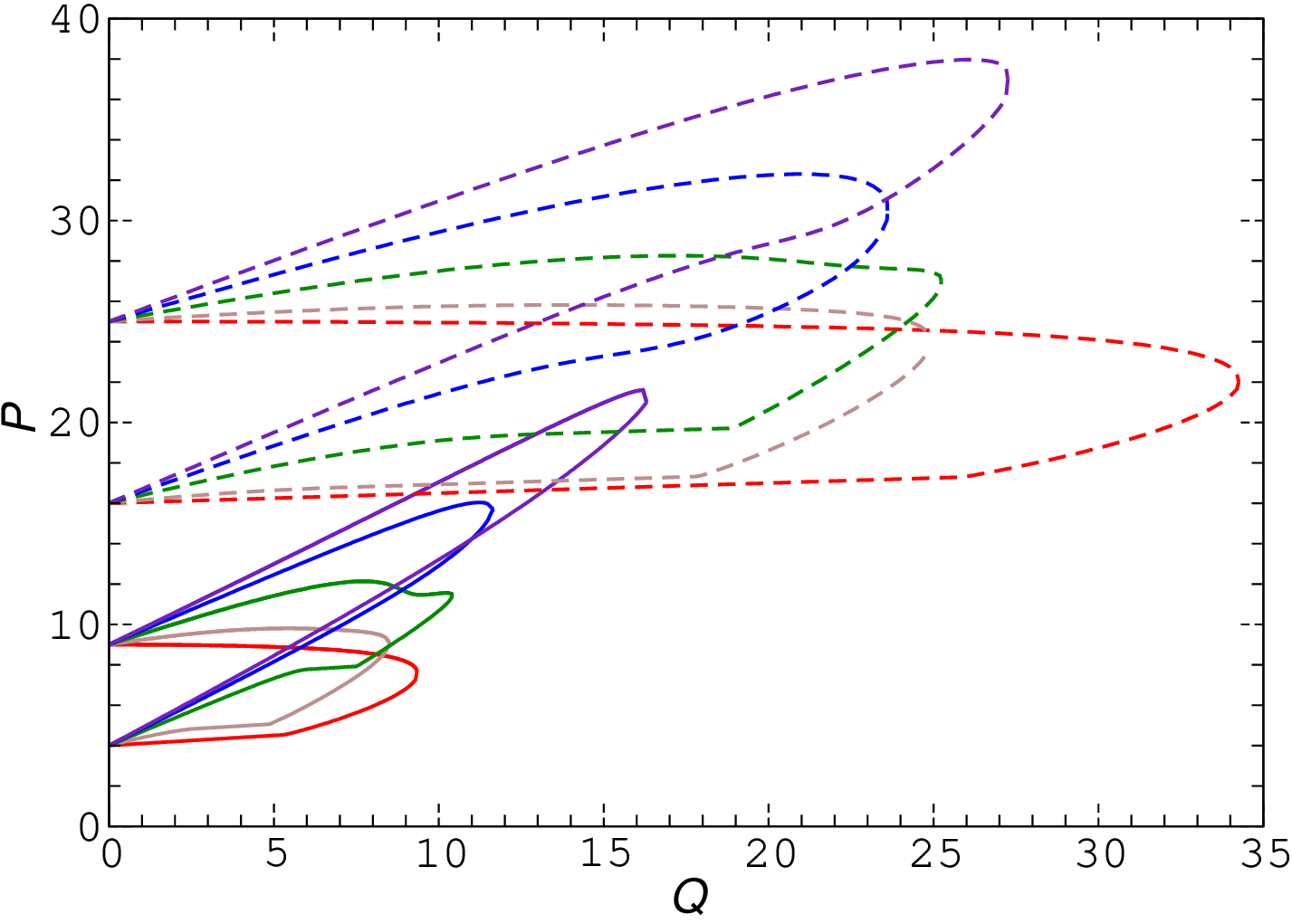}
 \caption{Higher real eigennumbers of Eq.~(34) at different $\alpha$. } 
 \end{figure}
 \begin{figure}
 \includegraphics[width=85mm]{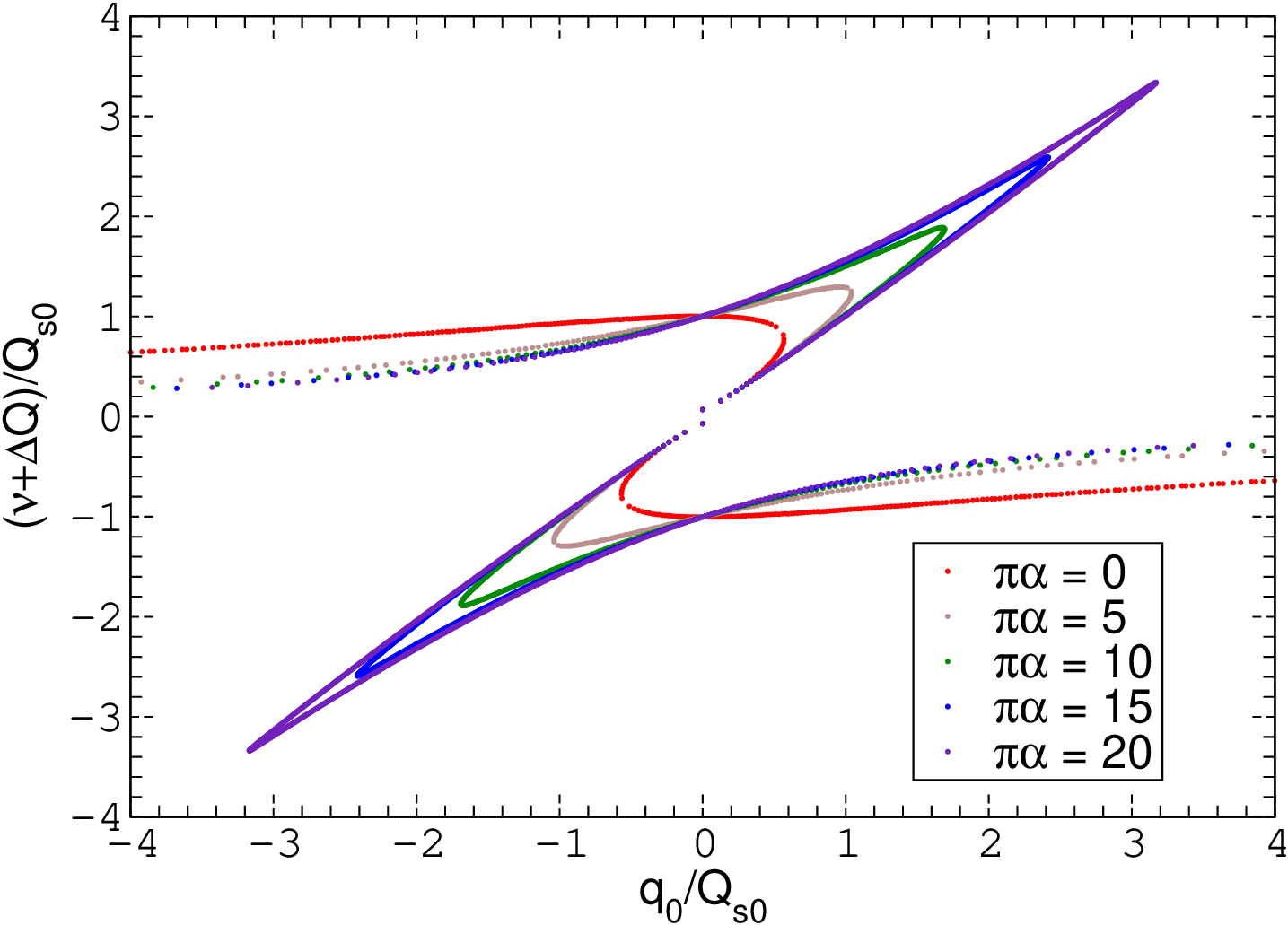}
 \caption{Lowest eigentunes of the bunch against the exponential wake 
 strength at $\,\Delta Q=0$.} 
 \end{figure}
The bunch eigentunes can be obtained at any SC tune shift with help of 
 Eq.~(27) applied to corresponding curve of Fig.~9 or Fig.~10.
The lowest eigentunes are represented in Figs. 11 and 12.
First of them demonstrates strong increase of the threshold when the damping 
 coefficient increases at $\,\Delta Q=0$. 
However, the dependence becomes weaker at higher SC tune shift as it is 
 illustrated by Fig.~12 at $\,\Delta Q/Q_{s0}=4$. 
There are additional data in Table II where threshold of this mode is 
 represented in the interval $\,\Delta Q/Q_{s0}\le 6$. 
It is seen that, at  $\,\Delta Q/Q_{s0}>\sim 5$, the threshold goes down at 
 higher $\alpha$.

Behavior of the higher modes is not so much dependent on $\,\alpha$, 
 and is rather well illustrated by green and blue lines in Fig.~5.
The mode produced by the coalescence of the multipoles $\,m=-1$ and $-2$ 
 becomes the most unstable at $\,\Delta Q/Q_{s0}>5-10$, dependent on 
 $\,\alpha$.
The general picture is shown in Fig.~13, and it demonstrates that the 
 dependence of the threshold on the wake strength is almost linear at 
 $\,\Delta Q/Q_{s0}>~10$. 

\begin{figure}
\includegraphics[width=85mm]{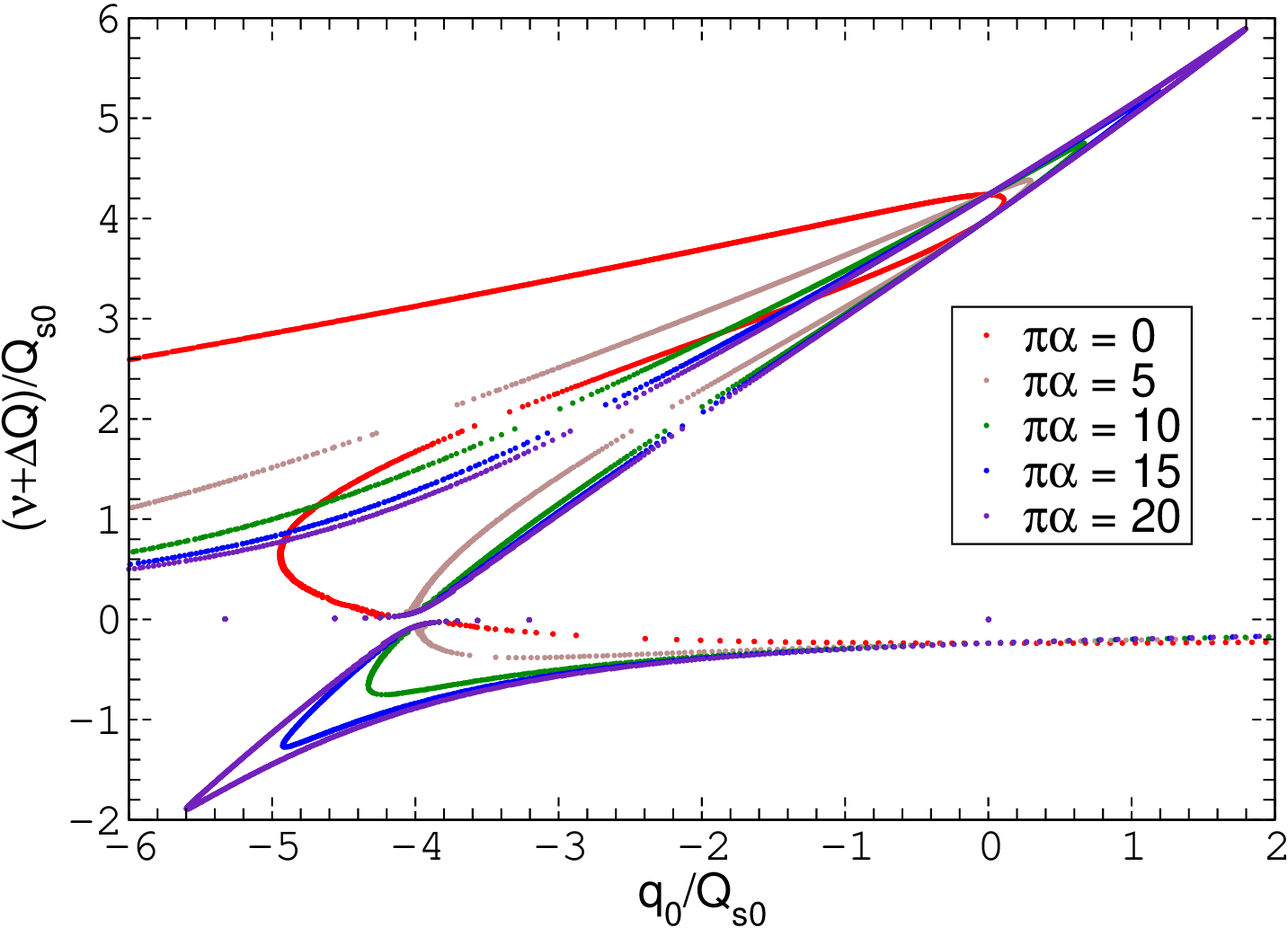}
\caption{Lowest eigentunes of the bunch against the exponential wake 
 strength at $\,\Delta Q=4$.} 
\vspace{10mm}
\end{figure}
\begin{table}
\caption{The TMCI threshold of the exponential wake due to a coalescence of 
 the lowest multipoles $\,m=0$ and $\,m=-1$ at modest space charge tine shift.}
\vspace{10mm}
\begin{tabular}{|l|c|c|c|c|c|c|c|}
\hline 
$\Delta Q/Q_{s0}\rightarrow$ &~0~&~1~&~2~&~3~&~4~&~5~&~6~\\
\hline
$\pi\alpha=0$  &--0.57~&--1.10~&--2.00~&--3.30~&--4.95~&--6.95~&--9.29~\\
$\pi\alpha=5$  &--1.03~&--1.47~&--2.23~&--3.04~&--4.00~&--5.08~&--6.20~\\
$\pi\alpha=10$&--1.71~&--2.19~&--2.80~&--3.55~&--4.34~&--5.18~&--6.07~\\
$\pi\alpha=15$&--2.42~&--2.91~&--3.52~&--4.18~&--4.82~&--5.40~&--6.03~\\
$\pi\alpha=20$&--3.20~&--3.69~&--4.28~&--4.90~&--5.45~&--6.00~&--6.60~\\
\hline
\end{tabular}
\end{table}

Asymptotic behavior of the threshold can be obtained by plotting of the 
 tangent to the curves in Fig. 10, as it has been explained at the end of 
 Sec.~III.
The asymptotic formula is
\begin{eqnarray}
  (q_0)_{tresh}=k(\alpha)\Delta Q
\end{eqnarray}
 with coefficients $\,k(\alpha)$ given by Fig~14. 
According to this plot, dependence of the TMCI threshold on the damping 
 factor is not very strong being almost constant at $\,\pi\alpha\ge 12$.
It happens due to normalization which has been used in Eq.~(34) to reach 
 $\,\nu_{rigid}=q_0$.
Much stronger dependence has been obtained in Ref.~\cite{Bl1} because the 
 relation $\,\kappa=2$ has been actually used there.
With this correction, the results are rather close at $\,\Delta Q/Q_{s0}<20$.
 \begin{figure}
 \includegraphics[width=85mm]{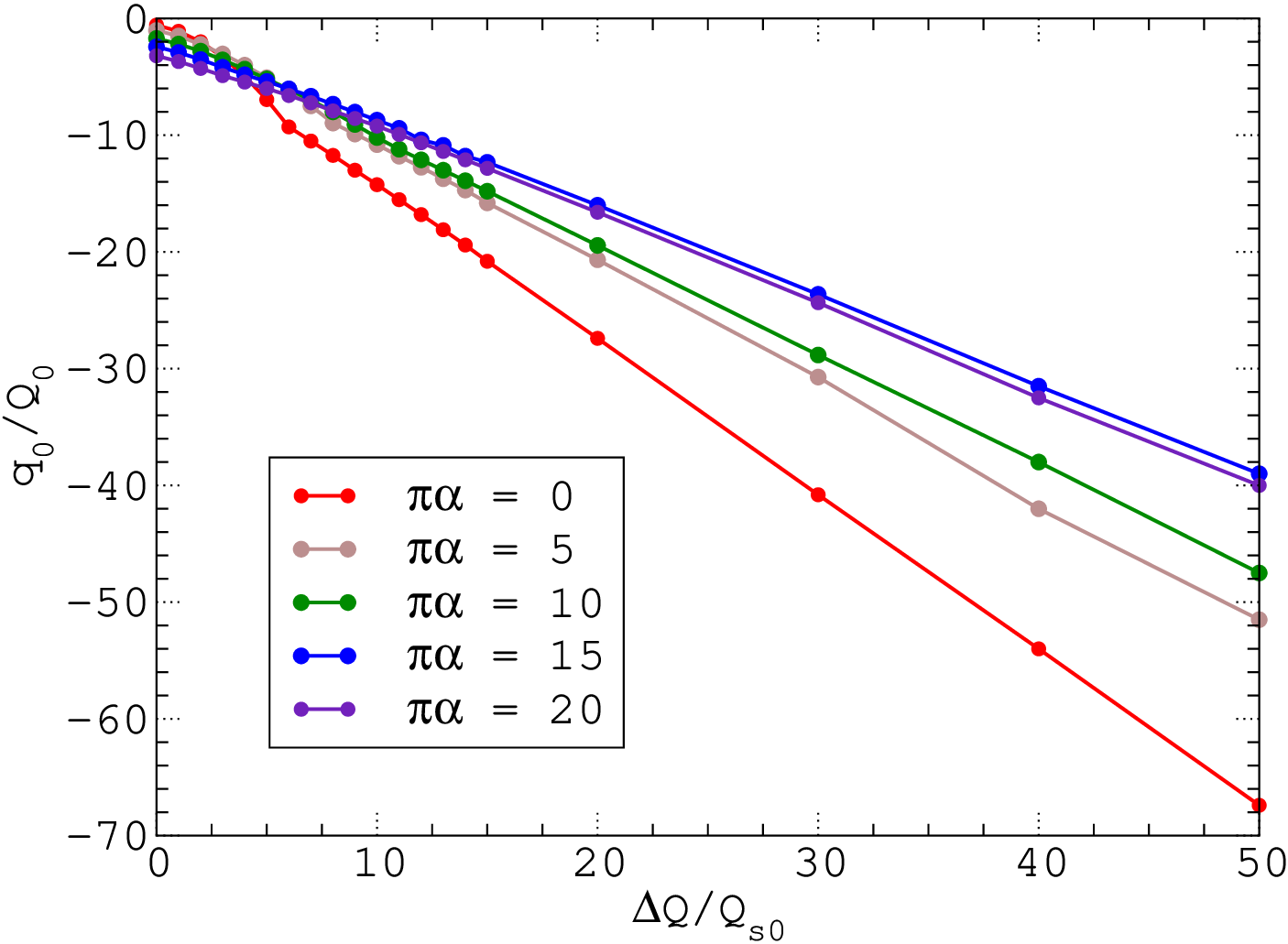}
 \caption{Instability threshold of exponential wake with different 
 $\,\alpha$ against the SC tune shift.} 
 \end{figure}
 \begin{figure}
 \includegraphics[width=85mm]{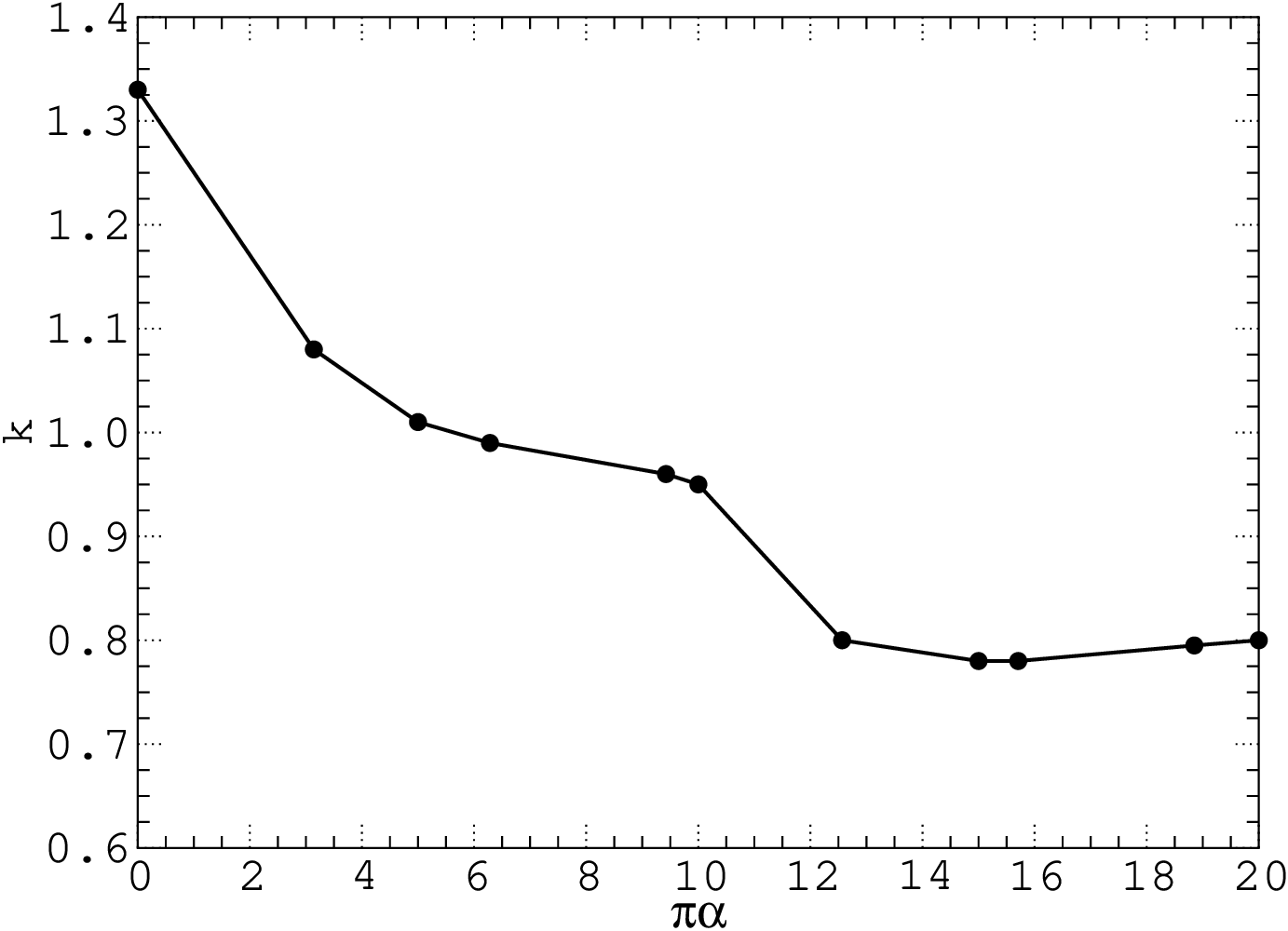}
 \caption{Asymptotic ratio $\,q_0/\Delta Q$ of exponential wake against 
 the damp coefficient.  } 
 \end{figure}

\section{Conclusions}

Transverse mode coupling instability is considered in the paper in frameworks 
 of the hollow bunch model in a square potential well with space charge tune 
 shift taken into account.
Two methods are used to calculate the instability threshold.

First of them is built upon the expansion technique using an infinite set of 
 basis functions with subsequent truncations of the series. 
Similar approximate method was used before by different authors to analyze 
 the bunch instability in a parabolic potential well, which circumstance 
 allows to compare the results. 
Received qualitative and quantitative resemlance enables us to extend the
 applicability of the square well model because it correctly describes the 
 lowest radial modes which are just responsible for the TMCI in parapolic 
 potential well.

However, the expansion technique is actually applicable only at a moderate 
 value of the space charge tune shift.
Therefore, another method is also offered and applied in the paper 
 consisting in a direct step-by-step solution of the integral-differential 
 equation for the bunch offset with any space charge.
It confirms the results of the expansion method in the area of its 
 applicability, and continues them to arbitrary large space charge.
The method is applied with the square, resistive, and various exponential 
 wake forms.
In all the cases, rather similar results are obtained for the normalized wake 
 amplitudes $\,q_0$ if the tune shift of the lowest (rigid) head-tail mode is 
 used each time as the scaling factor.
In particular, it is shown that the instability threshold is asymptotically 
 proportional to the tune shift: $\,(q_0)_{thresh}=k\Delta Q$ with the 
 coefficient $\,k=0.8-1.3$, dependent on the wake form.

The results allow to conclude that properly normalized square wake can be a 
 quite appropriate model for monotonous wake functions.
However, the problem remains open in the case of oscillating wake. 

\section{Acknowledgment}

Fermi National Accelerator Laboratory is operated by Fermi Research Alliance, LLC 
under Contract No. DEAC02-07CH11395 with the United States Department of Energy.




\begin{thebibliography}{99}
\bibitem{Bl1} M. Blaskiewicz, Fast head-tail instability with space charge, 
Phys. Rev. ST Accel. Beams 1, 044201 (1998).
\bibitem{Bu1} A. Burov,  Head-tail modes for strong space charge, 
Phys. Rev. ST Accel. Beams 12, 044202 (2009) and 12,109901 (2009).
\bibitem{Ba1} V. Balbekov,  Transverse instability of a bunched beam with 
space charge and wakefield, Phys. Rev. ST Accel. Beams 14, 094401 (2011).
\bibitem{Bl2} M. Blaskiewicz, Comparing new models of transverse instability 
with simulations, in  {\it Proceedings of the 3rd International Particle 
Accelerator Conference.} New Orleans, LA, 2012 (IEEE, Piscataway, NJ, (2012) 
\bibitem{Ba2} V. Balbekov, Single bunch transverse instability in a circular 
accelerator with chromaticty and space charge, JINST 10 P10032 (2015).
\bibitem{Ba3} V. Balbekov,  Transverse mode coupling instability with space 
charge and different wakefields,
Phys. Rev. Accel. Beams 20, 034401 (2017).
\bibitem{Ng1} B. Ng, Report No. Fermilab-FN-07-13 (2002).
\end{thebibliography}
\end{document}